\documentclass{svjour2}                    
\smartqed  
\usepackage{graphicx,comment}
\usepackage{color,ulem,amssymb,amsmath}
\definecolor{greencolor}{rgb}{0,0.5,0.2}
\definecolor{redcolor}{rgb}{.7,0.,0.}
\definecolor{bluecolor}{rgb}{0,0.,1.}
\definecolor{greycolor}{rgb}{.5,.5,.5}

\begin{document}

\title{Identifying trends in word frequency dynamics}
\titlerunning{Word frequency dynamics}     

\author{Eduardo~G.~Altmann   
\and Zakary~L.~Whichard
 \and  Adilson~E.~Motter}

\institute{E. G. Altmann \at
              Max Planck Institute for the Physics of Complex Systems,  01187 Dresden, Germany\\
              Tel.: +49-351-871-2415\\
              \email{edugalt@pks.mpg.de}          
           \and
           Z. L. Whichard \at
           Department of Physics and Astronomy, Northwestern University, Evanston, IL 60208, USA\\
            Tel.: +1-847-491-4602\\
              \email{z-whichard@northwestern.edu}  
           \and
           A. E. Motter \at
           Department of Physics and Astronomy and  Northwestern Institute on Complex Systems, Northwestern University, Evanston, IL 60208, USA\\
             Tel.: +1-847-491-4611\\
              \email{motter@northwestern.edu}  
}

\date{Received: date / Accepted: date}

\maketitle

\vspace{-0.5cm}
\begin{abstract} 
The word-stock of a language is a complex dynamical system in which words can be created, evolve, and become extinct.
Even more dynamic are the short-term fluctuations in word usage by individuals in a population.  Building on the recent demonstration that {\it word niche} is a strong determinant of future rise or fall in word frequency, here we introduce a model that allows us to distinguish persistent from temporary increases in frequency. Our model is illustrated using a  $10^8$-word database from an online discussion group  and a $10^{11}$-word collection of digitized books.  The model  reveals a strong relation between changes in word dissemination and changes in frequency. Aside from their implications for short-term word frequency dynamics, these observations are potentially important for language evolution as new words must survive in the short term in order to survive in the long term.
\end{abstract}

\section{Introduction}\label{intro}

Quantitative studies of natural languages have led to significant advances in the understanding of word statistics~\cite{Manning,Baayen2002} and language evolution~\cite{Pagel2009,Gell-Mann2011}. A comparatively less explored (albeit extremely important)  problem concerns the dynamics of word usage. Some representative examples include the study of bursts and lulls in word recurrence in online communities~\cite{PLoS1}, distributions of $n$-grams in books written over the past 200 years~\cite{Michel2010}, and analysis of word content in Twitter posts to assess temporal changes in perceived happiness~\cite{Dodds2011}. Language evolution and word statistics are related to word dynamics, as illustrated, for example, by early findings that word frequency itself is a correlate of word success at historical time scales~\cite{Lieberman2007,Pagel2007}. At shorter time scales, however, this relation is more subtle and remains far less understood.

For time scales of just a few years, we have recently shown that word niche is a stronger determinant of future change in word frequency
usage than the initial word frequency itself~\cite{PLoS2}.  The niche of a word was defined in terms of the number of people and topics
making use of the word and quantified by dissemination coefficients $D^{(\cdot)}$. These measures were applied to large records of
Usenet groups spanning approximately two decades, in which people are represented by Usenet users and topics are represented by the
discussion threads.  In particular, the results in~\cite{PLoS2} show: ({\it i}) that the dissemination across users, $D^U (t_1)$, and threads, $D^T(t_1)$, at a time $t_1$ are both strongly positively correlated with the change in $\log$-frequency $\Delta \log f_{t_2,t_1} =\log_{10}f(t_2) -\log_{10}f(t_1)$ for $t_2-t_1$ of a few years;  ({\it ii}) the changes in dissemination $\Delta D^U_{t_2,t_1} = D^U (t_2 ) - D^U (t_1)$ and $\Delta D^T_{t_2,t_1} = D^T (t_2 ) - D^T (t_1)$
are both negatively correlated with $\Delta \log f_{t_2,t_1}$ over the same time intervals.

\begin{figure}[t!] 
\begin{center}
\includegraphics[width=0.99\columnwidth]{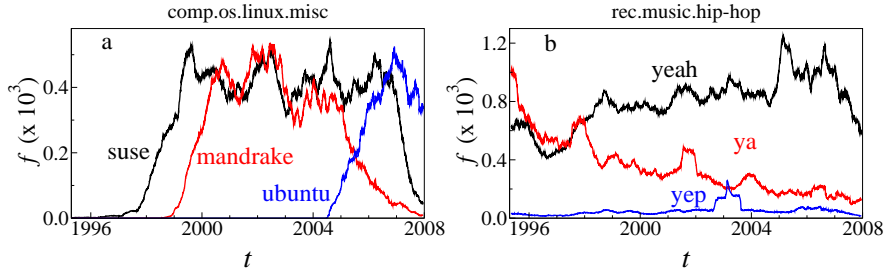}
\end{center}
\caption{Frequency dynamics for example words in the (a) Linux and (b) hip-hop groups. The frequency of a word is computed as the number of occurrences of  the word relative to the total number of words in a running window of half a year. }
\label{fig1}
\end{figure}

Here, we explore the relation between dissemination and frequency change using simple models for the population of word users.
We interpret our results  using data from two Usenet groups~\cite{data}: the comp.os.linux.misc group, which is focused on Linux operating systems and has 28,903 users and 140,517 threads for the period 1993-08-12 through 2008-03-31, and the rec.music.hip-hop group, which is focused on hip-hop music and has 37,779 users and 94,074 threads for the period 1995-02-08 through 2008-03-31. In these datasets, each post represents a unit of text and is associated with a user and thread, while each thread itself is defined by the initial post and all replies.  
Examples of the variation of word frequency  in these datasets are shown in Fig.~\ref{fig1}. Using our model and analysis of these datasets, we show that increase in frequency not accompanied by concurrent increase in the number of users  is reflected as a decrease in $D^U$ and subsequent frequency fall. 
This, along with the observations ({\it i}) and ({\it ii}), 
illuminates the mechanistic difference between temporary and persistent frequency changes
and helps explain why most frequency rises are just transient.  We focus on modeling $D^U$, with the view that analogous results hold for  $D^T$.

We also explore signatures of this behavior over longer time scales by considering a digitized collection of over 2.4 million  books published in English between 1820 and 2000~\cite{google_books}. In this case, the dissemination is considered across different books, which captures characteristics of both word users and topics. This dataset allows us to demonstrate that our observations are not unique to informal, Internet-based communications, and that they do in fact concern properties inherent to language change in general.

We believe these results are timely as numerous studies  are being carried out on
statistical physics aspects of natural languages. Such studies have considered properties on scales ranging from individual letters~\cite{Stephens2010} to thousands~\cite{Montemurro2010} or even millions~\cite{PLoS1} of words, and often benefit from concepts such as phase transitions~\cite{scaling_pt1,scaling_pt2} and techniques such as network representation~\cite{netw1,netw2,netw3,netw4}. In this context, increasing attention has been given to 
the modeling of language usage and language change (see, e.g.,~\cite{PLoS1,Serrano2009,Corral2009,Sole2010,Petersen2012,Perc2012}). 
Our study of factors distinguishing persistent from temporary word frequency change contributes to this growing body of literature.

\section{Results}

\subsection{Dissemination Coefficient}

We define the coefficient of dissemination of each word~$w$  across users as
\begin{equation}\label{eq.dissemination}
D^U_w=\frac{U_w}{\tilde{U}(N_w)}, 
\end{equation}
where $N_w$ is the number of occurrences of the word in the dataset, 
$U_w$ is the number of users whose posts include word~$w$ at least once,
and~$\tilde{U}$ is the expected number of users predicted by a baseline model in 
which words are randomized across users and threads.

Specifically, the baseline is defined from $\tilde{U}=\sum^{N_U}_{i=1}\tilde{U}_i$, where $N_U$ is 
the total number of users and $\tilde{U}_i$  is the probability that user $i$ would use word $w$ at least once 
if all words in the dataset are shuffled randomly while keeping fixed the sizes of the posts. 
The probability~$\tilde{U}_i$ can be calculated as the complement of the probability that the user never uses the word:
$
\tilde{U}_i=1-\prod_{j=0}^{N_w-1}\left[1-\frac{m_i}{N_A-j}\right], 
$
where $N_w$ be the number of occurrences of the word~$w$, $m_i$ be the total number of words 
contributed by user~$i$, and $N_A\equiv\sum_w N_w=\sum_i m_i$ is the total number of words in the dataset. 
In our datasets, $m_i/N_A\ll1$ and~$f_w\equiv N_w/N_A\ll1$, which allows us to further simplify this expression to
$
\tilde{U}_i \approx 1-e^{-f_w m_i}.
$
This represents a Poissonian baseline model in which the probability of using word $w$ is given by
the observed word frequency~$f_w$. For the rest of the paper, we drop the index $w$ for simplicity.

Therefore, the expected value of $D^U$ is $1$ for a word that is distributed randomly 
across all users. The main purpose of introducing this measure is to detect deviations 
from random. In particular,  $D^U<1$ represents words that are clumped and hence used 
above average by a subset of all users. For example, the word ``yep'' shown in Fig.~\ref{fig1}(b)
 has $D^U$ varying between $0.36$ and $0.90$ over different half-year windows. Clumping is in fact 
observed for most words in our datasets  (89\% of the words in the Linux group and 90\% of the 
words in the hip-hop group).
On the other hand,  $D^U>1$ represents words that are over disseminated and hence more 
evenly distributed across users than expected by chance. Greetings and expressions of 
gratitude,  such as {\it thanks}, 
tend to be in this class.
We refer to~\cite{PLoS2} for more information about the distribution of $D^U$ for the Usenet datasets we consider. 

\subsection{Statistical Model}  

We discuss a class of models that offer insights into how changes in $f$ are related to changes in $D^U$. 
This relation is key in discriminating between persistent and temporary word 
frequency growth.

Assume that each user $i$ and word $w$ are characterized by two quantities: $m_i\in[0,\infty)$, which is the size of the user's total 
contribution to the text in number of words, and $\nu_i\in[0,1]$, which is a fixed probability of using the word $w$ as opposed 
to any different word.  To simplify the calculations, we assume $m$ to be a continuous variable.
For each given word, a population of 
large size $N_U$
is then described by the joint probability density function $\rho(m,\nu)$ from which the relevant observable quantities can be
calculated.  In particular, within this model, the frequency of $w$ is given by 
\begin{equation}\label{eq.f}
f=\frac{N_w}{N_A}=\frac{\int_0^\infty dm\int_0^1 d\nu\,\, m \nu \rho(m,\nu)}{\int_0^\infty dm\,\, m\rho_m(m) },
\end{equation}
where~$\rho_m(m)\equiv \int_0^1 d\nu\,\, \rho(m,\nu)$. 
Moreover, the expected fraction of users of word~$w$  is 
$$
 \frac{U}{N_U}=1-\int_0^\infty  dm  \int_0^1d\nu \,\, \rho(m,\nu)e^{-m\nu},  
$$
and the baseline is
$$
 \frac{\tilde{U}}{N_U}=\int_0^\infty dm \int_0^1 d\nu \,\, (1-e^{-f m})\rho(m,\nu)=1-(\mathcal{L}\rho_m)(f),
$$
where the last term indicates the Laplace transform $(\mathcal{L}g)(y)\equiv\int_0^\infty dx\,\, g(x)
e^{-xy}$.  It follows from the ratio between the previous two equations that the dissemination $D^U$ is given by
\begin{equation}\label{eq.Dm}
D^U=\frac{1-\int_0^\infty dm   \int_0^1  d\nu \,\,  \rho(m,\nu)e^{-m\nu}}{1-(\mathcal{L}\rho_m)(f)},
\end{equation}
where $f$ is given by Eq.~(\ref{eq.f}). 

Therefore, given  a probability distribution $\rho(m,\nu)$,  Eq.~(\ref{eq.Dm}) provides a quantitative relation between frequency and dissemination.  As we proceed to our analysis of pertinent implications, we note that the main assumption involved in this derivation is
that users behave independently. That is, the size of their contributions as well as their individual word frequencies are independent
of those of the other  users. Nevertheless, this description is still quite general as it allows for an arbitrary relation between $m$ and $\nu$.

\subsection{Examples}

{\it Example 1:}  Assume that with respect to a word $w$  each user belongs to one of two distinct groups.
In the first group, formed by a fraction  $0\le q\le 1$ of the population, the users use the word with fixed frequency~$\nu=\nu^*$.
In the second group,  formed by the complementary fraction $1-q$ of individuals,  the users  use  the word with a negligible 
frequency ($\nu=0^+$). For simplicity we consider that  
all users contribute the same amount to the text, say $m^*$ words. Under these conditions, we have
\begin{equation}\label{eq.rho-ex1}
\rho(m,\nu)=\delta(m-m^*)[q\delta(\nu-\nu^*)+(1-q)\delta(\nu - 0^+)].
\end{equation}
In this case, Eq.~(\ref{eq.f})  results in the simple relation
\begin{equation}
f=\nu^* q
\label{eq.fnip}
\end{equation}
and Eq.~(\ref{eq.Dm}) leads to
\begin{equation}\label{eq.Dnip}
D^U = q \frac{1-e^{-m^*\nu^*}}{1-e^{-m^*\nu^* q}},
\end{equation}
where the term $m^* \nu^*$ corresponds to the average number of times each user uses the word $w$. 

\begin{figure}[h] 
\begin{center}
\includegraphics[width=0.99\columnwidth]{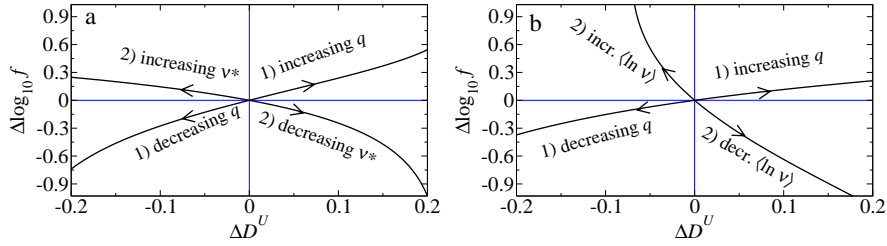} 
\end{center}
\caption{Frequency change and dissemination change for the statistical model.
(a) {\it Example~1}:~The changes  $\Delta \log f_{t_2,t_1}$ and $\Delta  D^U_{t_2,t_1}$
are determined using Eqs.~(\ref{eq.fnip}) and (\ref{eq.Dnip})  for $m^*=100$ words.
Starting with $q=0.5$ and $\nu^*= \nu^*_{1}\equiv 0.015$ at time $t_1$ (corresponding to the origin in the diagram), 
two scenarios are considered at time $t_2$: 
1)~$\nu^*=\nu^*_{1}$ and $0<q<1$  (curve in top right and bottom left quadrants); 
2)~$q=0.5$ and $0<\nu^*<1$   (curve in top left and bottom right quadrants).
 (b)  {\it Example~2}:~Same as in panel (a) but now using Eqs.~(\ref{eq.find}) and (\ref{eq.Dmi}), for  $\delta(\nu-\nu^*)$ replaced by a log-normal distribution with $\sigma_{\ln \nu} = 0.8$ and tunable $\langle \ln \nu \rangle$ and for $\delta(m-m^*)$ replaced by a log-normal distribution with $\sigma_{\ln m}= 1.36$ and  $\langle \ln m \rangle=4.9$. The first scenario is implemented using $\langle \ln \nu \rangle=-4.9$ and $0<q<1$, while the second is implemented using $q=0.5$ and $-10<\langle \ln \nu \rangle<0$.
Note that these scenarios represent respectively positive and negative correlations between frequency and dissemination changes.  
} 
\label{fig2}
\end{figure}

Word usage changes over time not only in frequency but also in dissemination. While the frequency in Eq.~(\ref{eq.fnip}) grows linearly with both $\nu^*$ and $q$, the dissemination  coefficient in Eq.~(\ref{eq.Dnip}) increases with~$q$ but decreases with~$m^* \nu^*$. To understand the significance of this, we examine the two different scenarios  shown in Fig.~\ref{fig2}(a). In the first 
scenario, 
the frequency $\nu^*$ remains fixed but the fraction $q$ of the population using the word changes over time; for increasing $q$, this represents a situation in which the overall frequency $f$ increases because the word is used by more individuals. In the second scenario, the frequency $\nu^*$ changes, while the fraction $q$ of users of the word remains fixed;  
for increasing $\nu^*$, this corresponds to a case in which the frequency $f$ rises simply because the word is used more repetitively by the same individuals. It is then clear that an increase in either $\nu^*$ or $q$ leads to an increase in the overall frequency ($\Delta \log f>0$),  but increase in $\nu^*$ without a concurrent increase in $q$ leads to a decrease in dissemination ($\Delta D^U <0$) even though the number of adopters of the word does not decrease. On the other hand, an increase in $q$, and hence in the number of actual users of the word, causes both frequency and dissemination to increase. Given that $D^U(t_1)$ is strongly positively correlated with $\Delta \log f_{t_2,t_1}$~\cite{PLoS2}, it is clear that the first 
 scenario
 may lead to sustainable growth in frequency while the second may not. 

These conclusions do not depend sensitively on the assumption that the users contribute the same amount to the text. For example,  replacing Eq.~(\ref{eq.rho-ex1}) with 
$
\rho(m,\nu)=\rho_m(m)[q\delta(\nu-\nu^*)+(1-q)\delta(\nu-0^+)]
$
leads to the same relation for the frequency and to a slightly less explicit expression for the dissemination,
$$
D^U = q\frac{1-\int_0^\infty dm\,\,  \rho_m (m) e^{-m\nu^*}}{1-\int_0^\infty dm\,\, \rho_m(m) e^{-m\nu^*q }},
$$
which is qualitatively similar to Eq.~(\ref{eq.Dnip}) if the distribution $\rho_m(m)$ is peaked around a certain average $m^*$.

\bigskip
\noindent
{\it Example 2:}  In the  example above the variables $m$ and $\nu$ are assumed to be independent, i.e., $\rho(m,\nu)=\rho_m(m)\rho_\nu(\nu)$, meaning that the
probability of using the word~$w$ is independent of the size of the contribution of the
user. More generally, this case leads to
\begin{equation}\label{eq.find}
f=\int_0^1 d\nu \,\, \nu \rho_\nu(\nu), 
\end{equation}
and
\begin{equation}\label{eq.Dmi}
D^U=\frac{1-\int_0^1 d\nu \,\, \rho_\nu(\nu)(\mathcal{L}\rho_m)(\nu)}{1-(\mathcal{L}\rho_m)(f)}.  
\end{equation}
We have previously observed that $\rho_m(m)$ 
follows a log-normal distribution for the datasets considered here~\cite{PLoS2}.
In addition, by considering words of sufficiently high frequency to generate reliable statistics, we suggest that
 $\rho_{\nu}(\nu)\arrowvert_{\nu>0}$  too can be approximated by a  log-normal distribution.
Therefore, we consider the case in which $\ln m$ is a normal distribution with average  $\langle \ln m \rangle$ and standard deviation  $\sigma_{\ln m}$ for the whole population, and $\ln \nu$ is a normal distribution with average $\langle \ln \nu \rangle$ and standard deviation $\sigma_{\ln \nu}$ for a fraction $q$ of 
the population. 
Here, $q$ represents the fraction of users with $\nu>0$ and hence a non-negligible probability of using the word under consideration;  the remaining fraction $1-q$ of users do not use the word and are assigned $\nu=0^+$. 

Figure~\ref{fig2}(b) shows a realization of this model for a choice of parameters representative of those in our datasets. Like in the case of the previous example, in the scenario in which the number of users is varied by controlling $q$, the resulting changes in the overall frequency  are accompanied by concordant changes in dissemination  ($\Delta \log f \times \Delta D^U >0$); conversely, the scenario in which the frequency is varied for a fixed number of users (now by controlling $\langle \ln \nu \rangle$), the changes in the overall frequency are accompanied by opposing changes in dissemination ($\Delta \log f \times \Delta D^U <0$). As already mentioned, owing to the positive correlations between dissemination and subsequent  frequency changes~\cite{PLoS2}, the first of these two scenarios will generally lead to more sustainable changes in frequency. This implication is demonstrated explicitly in the next section.

\subsection{Empirical Observations}

To test the behavior of words in real datasets, we performed additional analysis in the Linux and hip-hop Usenet groups~\cite{data}. Motivated
by Fig.~\ref{fig2}, we focus on concurrent changes of both $\log f$ and $D^U$.  Specifically, we measured  $\Delta \log f_{t_2, t_1}$ and $\Delta D^U_{t_2, t_1}$ for non-overlapping half-year windows centered at times $t_1$ and $t_2 = t_1+\Delta t$ years.  
We consider all words in the intermediate frequency range $10^{-7} \lessapprox f < 3\times 10^{-4}$,  for which $D^U$ has been observed not to depend strongly on $f$. 
This independence facilitates analysis of the separate  influence of frequency and dissemination on frequency change.
In order to avoid floor effects on extremely low-frequency words and ceiling effects on extremely high-frequency ones, this was implemented by only selecting words that appear more than $5$ times in both windows and with a frequency no larger than $3\times 10^{-4}$ in any window.

In our analysis, words are strings composed only by the symbols $``a-z,',-''$   
and are subjected to no additional lemmatization (we refer to~\cite{PLoS2} for the filtering of spams in our datasets).  Taking all windows into account, $32,795$ different unique words passed these criteria for the Linux group and $27,869$ for the hip-hop group, 
corresponding to more than $40\%$ of the whole text in each case; the whole text consists of $7.2\times10^7$ and $5.3\times10^7$ word 
occurrences, respectively. 

\begin{figure} 
\begin{center}
\includegraphics[width=0.99\columnwidth]{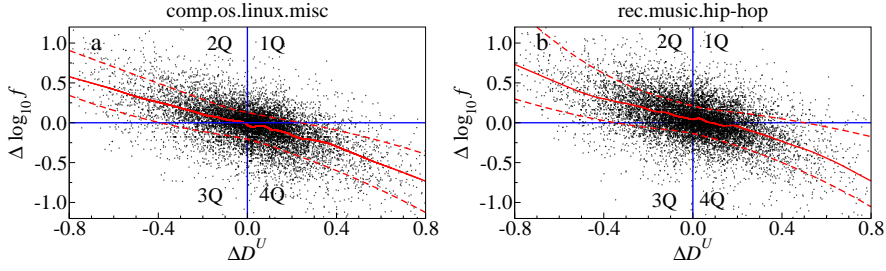}
\end{center}
\caption{Frequency change versus dissemination change for the (a) Linux and (b) hip-hop groups.
Both $\Delta D^U_{t_2, t_1}$ and $\Delta \log f_{t_2, t_1}$ are calculated over half-year windows separated by two years, and centered on $t_1=$ 1998-01-01 and  $t_2=$ 2000-01-01.
The scatter plots include all words with $N_w>5$ in both windows,
whereas the continuous 
lines indicate the running medians
and the dashed lines indicate the 5th and 95th running percentiles.
Words with rising frequency appear above and words with falling frequency appear below $\Delta \log f_{t_2, t_1}=0$. The higher concentration of points in the second and fourth quadrants indicate that frequency increase (decrease)  is for most words accompanied by dissemination decrease (increase), which corresponds to scenario 2 in Fig.~\ref{fig2}.
}
\label{fig3}
\end{figure}

Figure~\ref{fig3} shows $\Delta \log f$ and $\Delta D^U$ for $t_1=$ 1998-01-01 and $\Delta t=2$ years. The distribution of words in each scatter plot is centered around the origin and spread over all quadrants. However, the distribution is clearly biassed towards the second and fourth quadrants. This is a manifestation of the negative correlations that dominate the relation between frequency change and dissemination change. The tendency of $\Delta \log f$ and $\Delta D^U$ to vary in opposite directions  is evident also from the running median in $\Delta \log f$ as a function of $\Delta D^U$ (Fig.~\ref{fig3}, continuous 
lines). In view of the properties of the statistical model in Fig.~\ref{fig2}, this indicates that, for most words exhibiting a significant variation in overall frequency, the observed variation occurs due to a change in the usage rate among existing users of the word rather than a change in the number of individuals adopting the word. 

\begin{figure}
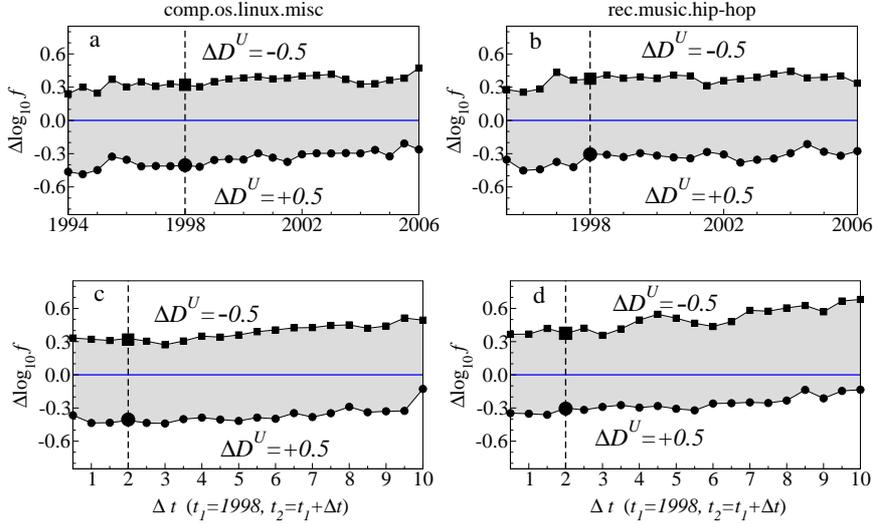

\begin{center}
\includegraphics[width=0.99\columnwidth]{Fig4ab.eps}

\vspace{0.1cm}  
\includegraphics[width=0.97\columnwidth]{Fig4cd.eps}
\vspace{-0.2cm}
\end{center}
\caption{Pattern of frequency change as a function of time for the (a,~c) Linux and (b,~d) hip-hop groups. 
(a,~b) Medians of the frequency change $\Delta \log f_{t_2, t_1}$ as a function of the time $t_1$ for given $\Delta D^U_{t_2, t_1}$ between $-0.5$ (solid squares) and $0.5$ (solid circles);
the windows are half-year wide and centered at $t_1$, and $t_2=t_1+ 2$ years.
(c,~d) Medians of the frequency change $\Delta \log f_{t_2, t_1}$ as a function of the time interval $\Delta t=t_2- t_1$ for given $\Delta D^U_{t_2, t_1}$  between  $-0.5$ (solid squares) and $0.5$ (solid circles); the windows are half-year wide and centered on $t_1=$ 1998-01-01, and $t_2=t_1+ \Delta t$ years.
In all panels, we consider all non-overlapping windows and the emphasized symbols correspond to the window pair in Fig.~\ref{fig3}. 
The word selection is the same used in Fig.~\ref{fig3}.
}
\label{fig4}
\end{figure}

To verify the generality of these observations, we consider  the values of the running median at $\Delta D^U=\pm 0.5$ as a quantitative indicator of the general relation between $\Delta \log f$ and $\Delta D^U$.  As shown in Fig.~\ref{fig4}(a, b), this indicator does not change substantially when we vary the position $t_1$ of the initial window. This implies that the conclusions drawn from Fig.~\ref{fig3} are in fact typical in our datasets for changes in frequency and dissemination over the time scale of a few years. Moreover, Fig.~\ref{fig4}(c, d) shows that similar robustness is also observed when we vary this time scale, represented by the time $\Delta t$ between the initial and final window. The values of the median of $\Delta \log f$ at $\Delta D^U=\pm 0.5$ increase slightly for large $\Delta t$, but this can be attributed in part to the criterion  $N_w> 5$, which has the effect of selecting against negative frequency changes and does so more strongly as the time between the windows  is increased. The distance between these values is therefore a more informative measure than the values themselves, and this measure does not change substantially with $\Delta t$. In all cases, the median of $\Delta \log f$ at $\Delta D^U= - 0.5$ is significantly larger than at $\Delta D^U= + 0.5$, confirming that large short-term variations in frequency and dissemination tend to oppose each other.  Nevertheless, for given $t_1$ and $\Delta t$, a significant number of individual words do exhibit variations in frequency and dissemination that are concurrently increasing or decreasing, as illustrated in Fig.~\ref{fig3}. 

\begin{figure}
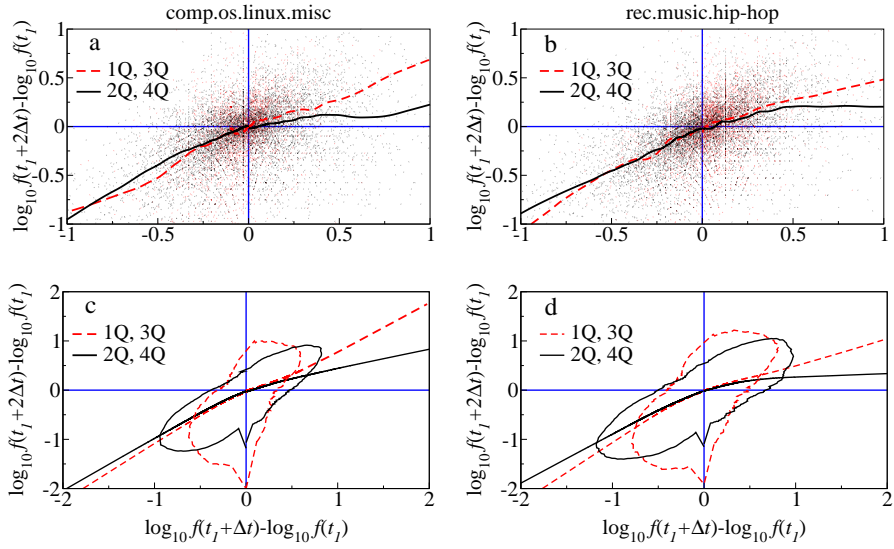

\begin{center}
\includegraphics[width=0.99\columnwidth]{Fig5ab.eps}
\includegraphics[width=1.00\columnwidth]{Fig5cd.eps}
\end{center}
\caption{Persistency of frequency change for the (a,~c) Linux and (b,~d) hip-hop groups. 
(a,~b) Frequency change $\Delta \log f_{t_1+2\Delta t, t_1}$ (after $2\Delta t$ years) versus  frequency change $\Delta \log f_{t_1+\Delta t, t_1}$ (after $\Delta t$ years) for $t_1=$ 1998-01-01 and $\Delta t= 2$ years; all three windows are half-year wide.
The dashed and continuous lines correspond to the running medians for points (shown in the background) 
with $\Delta \log f_{t_1+\Delta t, t_1}$  in the quadrants 1Q, 3Q and 2Q, 4Q of Fig.~\ref{fig3}, respectively.
(c,~d) Running medians as in (a, b) but now calculated using all points from all non-overlapping half-year windows for 
 $t_1$ ranging from 1994-01-01 to 2004-01-01 for the Linux group
 and from 1995-07-01 to 2004-01-01 for the  hip-hop group. 
The closed curves indicate the fraction of points along the corresponding directions from the origin. 
 The word selection is the same used in Fig.~\ref{fig3} except that,
 in order to keep all eligible words of the first two windows,
 the condition $N_w>5$ is not imposed in the third window. 
 }
\label{fig5}
\end{figure}

Finally, we demonstrate that frequency changes for which $\Delta \log f\times\Delta D^U>0$ are indeed more persistent than those for which $\Delta \log f\times\Delta D^U<0$. Figure~\ref{fig5}(a, b) illustrates this point by showing for $t_1=$ 1998-01-01 how a change in $\log f$ acquired over $\Delta t=2$ years  sustains itself after $2$ more years according to the quadrant the word belongs to in the representation of Fig.~\ref{fig3}. The running medians (Fig.~\ref{fig5}(a, b), 
dotted and continuous lines) 
indicate that the words belonging to the first quadrant ($\Delta \log f>0$,  $\Delta D^U>0$) exhibit a larger increase in frequency after $\Delta t +2$ years than the words in the second quadrant ($\Delta \log f>0$, $\Delta D^U<0$).  Likewise, although to a smaller extent, the words belonging to the third quadrant ($\Delta \log f<0$,  $\Delta D^U<0$) tend to exhibit a larger final decrease in frequency than the words in the fourth quadrant  ($\Delta \log f<0$, $\Delta D^U>0$). As shown in Fig.~\ref{fig5}(c, d),  for both the Linux and the hip-hop datasets, these systematic differences are statistically significant and continue to exist when $t_1$ is varied. 

\subsection{Confirmation over Longer and Larger Scales}

We consider the
Google Books Ngram Corpuses
of English-language publications over the period 1820-2000, which includes a total of 2,424,241 books~\cite{google_books}.  Starting with the raw data, we performed an initial cleaning to remove non-words. We focused on words formed by any combination of letters, apostrophes, and internal hyphens, containing at least 3 letters and less than 50 characters. 
Within this dataset,  upper- and lower-case letters are treated as different words, but it can be argued that distinguishing case has little impact on our results. This leads to a dataset of $1.7 \times10^{11}$ words.  
Within this set, we study the dissemination properties of words with average frequency in the interval $10^{-8} < f <  10^{-4}$, which results in $6.8\times10^{10}$ words and 632,912 unique words. In calculations of the dissemination coefficient, we further limit ourselves to words with a frequency of at least $10^{-7}$ within the corresponding year, which implies at least 10 occurrences of each selected word even for the years with the smallest number of books.

We consider  the dissemination across books,  with the associated dissemination coefficient $D_{w}^B$ given by
\begin{equation}
D_{w}^B= \frac{B_{w}}{\tilde{B}(N_{w})},
\end{equation}
where the actual number of books using the word, $B_{w}$, and the expected number predicted by the baseline model  $\tilde{B}(N_{w})$, are defined and calculated analogously to $U_{w}$ and  $\tilde{U}(N_{w})$ in the user dissemination coefficient in Eq.~(\ref{eq.dissemination}). All calculations of the dissemination coefficient $D_{w}^B$ are performed over time windows of one year.
Because no information is available in the database about the length of individual books, in estimating $\tilde{B}(N_{w})$ we have approximated the length of the books by their average length.  We focus on books published no earlier than 1820 to avoid conflation of the now obsolete long ``s'' with ``f'', which were not distinguished in the digitization process. Our choice of  the period 
1820-2000 is further motivated by the need to avoid years with extremely small and extremely large number of digitized books.

\begin{figure}
\begin{center}
\includegraphics[width=0.99\columnwidth]{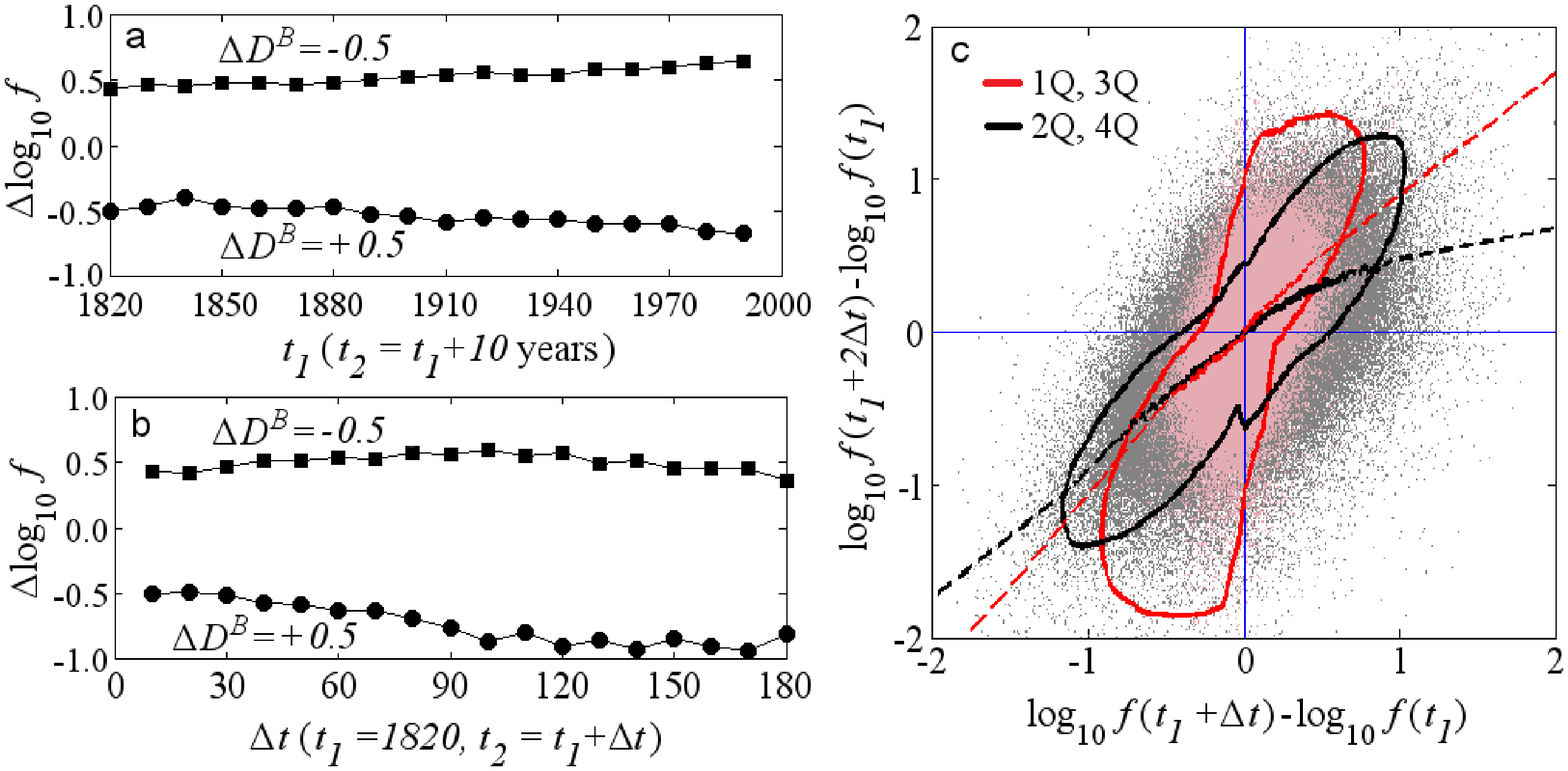}
\end{center}
\caption{Frequency change and dissemination change in the Google Books dataset. 
(a)  Medians of the frequency change $\Delta \log f_{t_2, t_1}$ as a function of the time $t_1$ for $\Delta D^B_{t_2, t_1}$ equal to $-0.5$ (solid squares) and $0.5$ (solid circles);
the windows are at $t_1$ and $t_2=t_1+ 10$ years.
(b) Medians of the frequency change $\Delta \log f_{t_2, t_1}$ as a function of the time interval $\Delta t=t_2- t_1$ for $\Delta D^B_{t_2, t_1}$  equal to  $-0.5$ (solid squares) and $0.5$ (solid circles); the windows are in $t_1= 1820$ and $t_2=t_1+ \Delta t$ years.
(c) Frequency change $\Delta \log f_{t_1+2\Delta t, t_1}$ (after $2\Delta t$ years) versus  frequency change $\Delta \log f_{t_1+\Delta t, t_1}$ (after $\Delta t$ years) for the aggregate collection of points corresponding to $t_1= 1820, 1830, ..., 1980$ and $\Delta t= 10$ years.
Points corresponding  to 
$\Delta \log f_{t_1+\Delta t, t_1}$ in the quadrants 1Q, 3Q and 2Q, 4Q of the $\Delta \log f$ versus $\Delta D^B$ plot (not shown)
are represented in red and black, respectively.  Following this color code, 
the closed curves indicate the fraction of points along the corresponding directions from the origin and
the dashed lines correspond to the running median for each quadrant. 
In all cases, the windows are one-year wide.
 }
\label{fig6}
\end{figure}

Figure~\ref{fig6} shows a summary of the empirical observations in this dataset. As in the case of the Usenet groups, the frequency change is negatively correlated with the dissemination change. This is illustrated both by considering a fixed $\Delta t=10$ years for $t_1$ varying from 1820 to 1990 (Fig.~\ref{fig6}(a)) and by considering a fixed $t_1=1820$ for $\Delta t$ varying from $10$ to $180$ years  (Fig.~\ref{fig6}(b)).  Over these long time scales, there are some systematic changes in $\Delta \log f$  both as a function of $t_1$ and as a function of $\Delta t$. But these changes may be partially due to the heterogeneity of the dataset. For example, because recent years have a larger number of books (and hence of words),  the smaller $\Delta \log f$ for $\Delta D^B=0.5$  for more recent $t_2$ may be in part due to the fact that statistical fluctuations are less likely to push infrequent words below the frequency threshold $10^{-7}$ in recent years than in early years. More important, even when we consider frequency change over relatively long time intervals, the sign of the 
accompanying change in dissemination is a determinant factor for subsequent changes in frequency.
This is illustrated in Fig.~\ref{fig6}(c) 
for frequency changes over 20 years  as determined by  the frequency and dissemination changes over the first 10 years. 
These empirical observations corroborate the conclusion that word dissemination plays a central role in future rise and fall of word frequency even over long times  and large social scales.

\section{Outlook}   

Our demonstration that {\it word frequency dynamics} can be statistically related to simple aspects of the {\it users' dynamics} opens new opportunities for the study of language dynamics in online  communities. 
Several aspects of language dynamics have been traditionally addressed by tacitly assuming a homogeneous and essentially 
passive population of users. This includes, for example, the long-term lexical evolution and its dependency on word frequency~\cite{Pagel2009}. This study, on the other hand, points to the importance of the medium in which the word is used, including its dynamics and heterogeneities, which determine the niche of the word~\cite{PLoS2}.  Our results clearly show that short-term frequency changes in which the increase (decrease) in frequency is accompanied by a concurrent increase (decrease) in dissemination are less dominant but far more persistent in a longer term.  

While we have focused 
mainly 
on the dissemination across users of a word, quantitatively described by the coefficient $D^U$, similar results hold for dissemination across topics ($D^T$), which is another important aspect of the word niche.  These different dimensions manifest themselves in the dissemination across documents in formal writing, which is both topic and author dependent, as observed in our analysis of the dissemination coefficient $D^B$ for digitized books. 
In online discussion groups and other informal settings,
because
word usage reflects one's social identity, it is likely that the words actually used by people depend more strongly on their social network than on the words they know. Future research may thus provide further insight into word usage dynamics by accounting for the possible influence of the underlying social network dynamics, and point to new directions within the growing body of literature on cognitively and socially informed models of language~\cite {Hruschka2009}.

Finally, we suggest that dissemination coefficients and the notion of niche itself can be extended to address factors contributing to success and failure in the spread of norms, 
propagation of information cascades, diffusion of innovation, and other processes that compete for adopters~\cite{RMP_2009}.
There are processes, such as the dynamics of fashions and  fads, in which an eventual widespread dissemination inhibits further adoption---a representative example being the selection of baby names~\cite{Kessler2012}. But because the initial adoption grows by imitation, even the rise of a fashion seems to depend critically on the positive feedback of dissemination~\cite{Zanette2012}. 
In  these  contexts, and in the dynamics of word usage too, another topic for future research concerns the impact of spatial patterns of dissemination (which is a major determinant in the survival of species and groups of species in ecological systems~\cite{Foote2008,Wilson2004,Meyer1996}) and their interactions with other dissemination measures.

\begin{acknowledgements}
We thank Janet Pierrehumbert for discussions during preliminary stages of the project.
This work was supported by the Northwestern University Institute on Complex Systems (E.G.A.), 
the Max Planck Institute for the Physics of Complex Systems (E.G.A.), and a Sloan Research Fellowship (A.E.M.).
\end{acknowledgements}

\newpage

~

\begin{thebibliography}{10}

\bibitem{Manning} 
Manning, C.D.,  Schuetze, H.: 
Foundations of Statistical Natural Language Processing.  
The MIT Press, Cambridge MA (1999)

\bibitem{Baayen2002}
Baayen, R.H.: 
Word Frequency Distributions. 
Springer, Berlin (2002) 

\bibitem{Pagel2009}
Pagel, M.:
Human language as a culturally transmitted replicator.
Nat. Rev. Genet. {\bf 10}, 405-415 (2009)

\bibitem{Gell-Mann2011}
Gell-Mann, M., Ruhlen, M.:
The origin and evolution of word order.
Proc. Natl Acad. Sci. {\bf 108}, 17290-17295 (2011)

\bibitem{PLoS1} 
Altmann, E.G.,  Pierrehumbert, J.B.,  Motter, A.E.:
Beyond word frequency: Bursts, lulls, and scaling in the temporal distributions of words.
PLoS ONE {\bf 4}(11),  e7678 (2009)

\bibitem{Michel2010}
Michel, J.-B. {\it et al.}:
Quantitative analysis of culture using millions of digitized books.
Science {\bf 331}, 176-182 (2010)

\bibitem{Dodds2011}
Dodds, P.S.,  Harris, K.D.,  Kloumann, I.M., Bliss, C.A., Danforth, C.M.:
Temporal patterns of happiness and information in a global social network: Hedonometrics and Twitter.
PLoS ONE {\bf 6}(12), e26752 (2011)

\bibitem{Lieberman2007}
Lieberman, E.,  Michel, J.-B.,  Jackson, J.,  Tang, T., Nowak, M.A.:
Quantifying the evolutionary dynamics of language.
Nature {\bf 449}, 713-716 (2007)

\bibitem{Pagel2007}
Pagel, M., Atkinson, A., Meade, A.: 
Frequency of word-use predicts rates of lexical evolution throughout Indo-European history.
Nature {\bf 449}, 717-720  (2007)

\bibitem{PLoS2} 
Altmann, E.G.,  Pierrehumbert, J.B.,  Motter, A.E.:
Niche as a determinant of word fate in online groups.
PLoS ONE {\bf 6}(5), e19009 (2011)

\bibitem{data}
The Usenet
Archives, available at http://groups.google.com

\bibitem{google_books}
The Google Books Ngram Corpuses, available at http://books.google.com/ ngrams/datasets


 \bibitem{Stephens2010} 
Stephens, G.J., Bialek, W.:
Statistical mechanics of letters in words.
Phys. Rev. E {\bf 81}, 066119 (2010) 

\bibitem{Montemurro2010} 
Montemurro, M., Zanette, D.H.:
Towards the quantification of the semantic information encoded in written language.
Adv. Compl. Sys. {\bf 13}, 135-153 (2010)
 
\bibitem{scaling_pt1} 
Ferrer i Cancho, R.,   Sol\'e., R.V.:
Least effort and the origins of scaling in human language.
Proc. Natl Acad. Sci. USA {\bf 100}, 788-791 (2003)

\bibitem{scaling_pt2} 
Prokopenko, M.,  Ay, N., Obst, O., Polani, D.:
Phase transitions in least-effort communications.
J. Stat. Mech. {\bf 2010}(11), P11025 (2010) 


\bibitem{netw1} 
Ferrer i Cancho, R.,  Sol\'e, R.V.:
The small world of human language.
Proc. R. Soc. Lond. B  {\bf 268}, 2261-2265 (2001)

\bibitem{netw2} 
Dorogovtsev, S.N., Mendes, J.F.F.:
Language as an evolving word web.
Proc. R. Soc. Lond. B  {\bf 268}, 2603-2606 (2001)
 
\bibitem{netw3} 
Motter, A.E, de Moura, A.P.S.,  Lai, Y.-C., Dasgupta, P.:
Topology of the conceptual network of language.
Phys. Rev. E {\bf 65}, 065102(R) (2002) 

\bibitem{netw4} 
Sigman, M., Cecchi, G.A.:
Global organization of the Wordnet lexicon.
Proc. Natl Acad. Sci. USA {\bf 99}, 1742-1747 (2002)

\bibitem{Serrano2009} 
Serrano, M.A., Flammini, A., Menczer, F.: 
Modeling statistical properties of written text. 
PLoS ONE {\bf 4}(4), e537 (2009)

\bibitem{Corral2009}  
Corral, R., Ferrer-i-Cancho, R., Boleda, G., Diaz-Guilera, A.:
Universal complex structures in written language. 
pre-print arXiv:physics.soc-ph/0901.2924v1 (2009)

\bibitem{Sole2010} 
Sol\'e, R.V.,  Corominas-Murtra, B., Fortuny, J.:
Diversity, competition, extinction: the ecophysics of language change.
J. R. Soc. Interface {\bf 7}, 1647-1664 (2010)

\bibitem{Petersen2012} 
Petersen, A.M., Tenenbaum, J.,  Havlin, S., Stanley, H.E.:
Statistical laws governing fluctuations in word use from word birth to word death.
Sci. Rep. {\bf 2},  313 (2012)

\bibitem{Perc2012} 
Perc, M.: 
Evolution of the most common English words and phrases over the centuries.
J. R. Soc. Interface {\bf 9}, 3323-3328 (2012)


\bibitem{Hruschka2009}  
Hruschka, D.J.,  Christiansen, M.H., Blythe, R.A., Croft, W., Heggarty, P., Mufwene, S.S., Pierrehumbert, J.B., Poplack, S.:
Building social cognitive models of language change.
Trends Cogn. Sci. {\bf 13}, 464-469 (2009)

\bibitem{RMP_2009}
Castellano, C.,  Fortunato, S.,   Loreto, V.:
Statistical physics of social dynamics.
Rev. Mod. Phys. {\bf 81}, 591-646 (2009)


\bibitem{Kessler2012}
Kessler, D.A., Maruvka, Y.E., Ouren, J., Shnerb, N.M.: 
You name it---How memory and delay govern first name dynamics. 
PLoS ONE {\bf 7}(6), e38790 (2012)

\bibitem{Zanette2012}
Zanette, D.H.:
Dynamics of fashion: The case of given names.
arXiv:1208.0576 [physics.soc-ph] (2012)

\bibitem{Foote2008}
Foote, M., Crampton, J.S., Beu, A.G., Cooper, R.A.: 
On the bidirectional
relationship between geographic range and taxonomic duration. Paleobiology
{\bf 34}, 421-433 (2008)

\bibitem{Wilson2004}
Wilson, R.J., Thomas, C.D.,  Fox, R., Roy, D.B.,  Kunin, W.E.: 
Spatial patterns in species distributions reveal biodiversity change.
Nature {\bf 432}, 393-396 (2004)

\bibitem{Meyer1996}
Meyer, M., Havlin, S., Bunde, A.:
Clustering of independently diffusing individuals by birth and death processes.
Phys. Rev. E {\bf 54}, 5567-5570 (1996)


\end{thebibliography}
\end{document}